\documentclass[
               twocolumn,
               prd,nofootinbib,superscriptaddress,
               showpacs,preprintnumbers,
               amsmath,amssymb]
              {revtex4}
\usepackage{graphicx,color}

\newcommand{\Tr}{\mathop\mathrm{Tr}\nolimits}

\newcommand{\Imag}{\mathop\mathrm{Im}\nolimits}
\newcommand{\Dm}{D_\mu}
\newcommand{\Dmu}{D^\mu}

\newcommand{\bra}[1]{\langle #1 |}
\newcommand{\ket}[1]{| #1 \rangle}
\newcommand{\bm}[1]{\boldsymbol{#1}}
\newcommand{\bk}{{\bm{k}}}
\newcommand{\bx}{{\bm{x}}}
\newcommand{\bkind}{^{\vphantom{*}}_\bk}

\newcommand{\fk}{f^{\vphantom{*}}_\bk}

\newcommand{\gsk}{g^{*}_\bk}
\newcommand{\half}{\frac{1}{2}}
\newcommand{\e}{\mathrm{e}}
\newcommand{\const}{\mathrm{const}}

\newcommand{\vpar}[2]{\frac{\delta #1}{\delta #2}}

\newcommand{\TeV}{\mathrm{TeV}}

\providecommand{\href}[2]{#2}

\begin{document}

\preprint{INR/TH-2003-6}

\title{Suppression of baryon number violation in electroweak collisions:
  Numerical results.}

\author{F.~Bezrukov}
\email{fedor@ms2.inr.ac.ru}
\affiliation{Institute for Nuclear Research of the Russian Academy of Sciences,\\
  60th October Anniversary prospect 7a, Moscow 117312, Russia}
\author{D.~Levkov}
\email{levkov@ms2.inr.ac.ru}
\affiliation{Institute for Nuclear Research of the Russian Academy of Sciences,\\
  60th October Anniversary prospect 7a, Moscow 117312, Russia}
\affiliation{Moscow State University, Department of Physics,\\
  Vorobjevy Gory, Moscow, 119899, Russian Federation}
\author{C.~Rebbi}
\email{rebbi@bu.edu}
\affiliation{Department of Physics---Boston University\\
  590 Commonwealth Avenue, Boston MA 02215, USA}
\author{V.~Rubakov}
\email{rubakov@ms2.inr.ac.ru}
\affiliation{Institute for Nuclear Research of the Russian Academy of Sciences,\\
  60th October Anniversary prospect 7a, Moscow 117312, Russia}
\author{P.~Tinyakov}
\email{Peter.Tinyakov@cern.ch}
\affiliation{Institute of Theoretical Physics, University of Lausanne,\\
  CH-1015 Lausanne, Switzerland}
\affiliation{Institute for Nuclear Research of the Russian Academy of Sciences,\\
  60th October Anniversary prospect 7a, Moscow 117312, Russia}

\date{May 27, 2003}

\begin{abstract}
  We present a semiclassical study of the suppression of
  topology changing, baryon number violating transitions induced by
  particle collisions in the electroweak theory.  We find that
  below the sphaleron energy the suppression exponent is remarkably 
  close to the analytic estimates based on a low energy expansion about the
  instanton.  Above the sphaleron energy, the relevant semiclassical
  solutions have qualitatively different properties from those
  below the sphaleron: they correspond to jumps on top of the barrier.  
  Yet these processes remain exponentially suppressed, and, 
  furthermore, the tunneling
  exponent flattens out in energy.  We also derive lower bounds on
  the tunneling exponent which show that baryon number violation
  remains exponentially suppressed up to very high energies of at
  least 30 sphaleron masses (250~TeV).
\end{abstract}



\maketitle

A long-standing problem in the electroweak theory is whether 
instanton-like processes occur at high rates in particle collisions
near and above the sphaleron energy.
The energy of the sphaleron~\cite{Klinkhamer:1984di} 
represents the minimum height of
the barrier separating topologically distinct vacua in a non-Abelian
gauge-Higgs theory, thus it sets a non-perturbative
energy scale at weak coupling.
Instanton-like transitions between
these vacua, which at low energies occur via tunneling and hence at
exponentially small rates, are energetically allowed to proceed
classically at energies above the sphaleron energy.  The problem is
whether or not these classical, and hence unsuppressed transitions are
allowed dynamically in collisions of highly energetic particles.  This
problem is particularly interesting in the context of the electroweak
theory, both because instanton-like transitions are accompanied by
non-conservation of baryon and lepton numbers~\cite{'tHooft:1976fv},
and because the sphaleron energy is relatively low,
$E_\mathrm{sph}\simeq8\TeV$.

As was first found in Refs.~\cite{Ringwald:1990ee,Espinosa:1990qn},
cross sections of collision-induced instanton processes 
increase rapidly with energy at $E\ll E_\mathrm{sph}$.  Subsequently,
it was
shown~\cite{McLerran:1990ab,Khlebnikov:1991ue,Yaffe:1990iy,Arnold:1990va}
that the total cross section has the exponential form\footnote{The
  subscript $HG$ here stands for ``holy grail'' \cite{Mattis:1992bj}.}
(for reviews see
Refs.~\cite{Mattis:1992bj,Tinyakov:1993dr,Rubakov:1996vz})
\begin{displaymath}
  \sigma_{tot} (E) \sim
  \exp \left\{
    -\frac{4\pi}{\alpha_W} F_{HG}(E/E_\mathrm{sph})
  \right\} \;,
\end{displaymath}
where $\alpha_W=g^2/4\pi$ is the small coupling constant
($\alpha_W\simeq1/30$ in the electroweak theory).  Perturbative
calculations about the instanton enable one to evaluate $F_{HG}$ as a
series in fractional powers of $E/E_\mathrm{sph}$, but the perturbative
expansion becomes
unreliable at $E\sim E_\mathrm{sph}$ and at higher energies.  Existing
analytical estimates of $F_{HG}$ at all
energies~\cite{Ringwald:2002sw,Ringwald:2003px}
are based on a number of assumptions which may or may not be valid.

One way to understand instanton-like processes at high energies
is to obtain numerically solutions to classical, real time field
equations exhibiting appropriate topology~\cite{Rebbi:1996zx}, and in
this way explore the region of parameter space where classical
over-barrier transitions do occur.  Besides the total energy $E$, 
an important
parameter is the number of incoming particles $N$, which one calculates
semiclassically for every solution.  This approach enables one to find
the approximate boundary of the classically \emph{allowed} region in the
$(E,N)$ plane; the analysis of Ref.~\cite{Rebbi:1996zx} extends to
$E\sim2E_\mathrm{sph}$ and shows that even at the highest energy
attained in this study the number of incoming particles is always large,
$N\simeq1/\alpha_W$, which is very far from realistic collisions.

In this paper we present the results of another computational
approach, which is appropriate for analyzing the classically
\emph{forbidden} region in the $(E,N)$ plane.  We study the
four-dimensional $SU(2)$ gauge theory with a Higgs doublet $\Phi$, which
corresponds to the bosonic sector of the Electroweak Theory
with $\theta_W=0$ and captures all relevant features of the Standard
Model (to leading order, the effects of fermions on the dynamics
of the gauge and
Higgs field can be ignored~\cite{Espinosa:1992vq}).  The
action of the model is
\begin{align}\label{SU2action}
  S = \frac{1}{4\pi\alpha_W}
      \int\! d^4x \left\{\vphantom{\frac{1}{2}}\right.
                         &
        -\frac{1}{2}\Tr F_{\mu\nu}F^{\mu\nu}
  \\
  & \left.\vphantom{\half}
      +(\Dm\Phi)^\dagger\Dmu\Phi
      -\lambda(\Phi^\dagger\Phi-4\pi\alpha_W v^2)^2
    \right\}
  \;. \nonumber
\end{align}
In most of our calculations the Higgs self-coupling $\lambda$ was set
equal to $\lambda=0.125$, which corresponds to $M_H=M_W$.  We found
that the dependence of our results on the Higgs boson mass is very weak, 
so the specific choice of $\lambda$ does not affect our conclusions.

Our starting point is the
observation~\cite{Rubakov:1992fb,Tinyakov:1992fn,Rubakov:1992ec} that
the inclusive probability of tunneling from a state with fixed energy
$E$ and fixed number of incoming particles $N$ is calculable in a
semiclassical way, provided that $E=\tilde{E}/\alpha_W$ and
$N=\tilde{N}/\alpha_W$, where $\alpha_W$ is a small parameter and
$\tilde{E}$ and $\tilde{N}$ are held fixed in the limit
$\alpha_W\to0$.  This inclusive probability is defined as follows,
\begin{equation*}
  \sigma(E,N) =
    \sum_{i,f} |\bra{f} \hat{S} \hat{P}_E \hat{P}_N \ket{i}|^2 \;,
\end{equation*}
where $\hat{S}$ is the $S$-matrix, $\hat{P}_{E,N}$ are projectors onto
subspaces of fixed energy $E$ and fixed number of particles $N$, and
the states $\ket{i}$ and $\ket{f}$ are perturbative excitations about
topologically distinct vacua.  In the regime $\alpha_W\to0$, with
$\tilde{E}$ and $\tilde{N}$ held fixed, this
probability can be calculated in the semiclassical approximation,
leading to
\begin{equation*}
  \sigma(E,N)\sim \exp\left\{
    -\frac{4\pi}{\alpha_W}F(\tilde{E},\tilde{N})
  \right\}
  \;,
\end{equation*}
where the exponent $F(\tilde{E},\tilde{N})$ is obtained by solving a
classical boundary value
problem~\cite{Rubakov:1992fb,Tinyakov:1992fn,Rubakov:1992ec} about
which we will have more to say later.

Furthermore, it has been
conjectured~\cite{Rubakov:1992fb,Tinyakov:1992fn,Rubakov:1992ec} that
the exponent for the two-particle cross section is recovered in the
limit of small number of incoming particles,
\begin{equation}\label{5*}
  F_{HG}(\tilde{E})=
  \lim\limits_{\tilde{N}\to 0}F(\tilde{E},\tilde{N})
  \;.
\end{equation}
This conjecture was checked in several orders of perturbation theory in
$E/E_\mathrm{sph}$ in gauge
theory~\cite{Tinyakov:1992fn,Mueller:1993sc} and by comparison
with the full quantum mechanical solution in a model with two degrees of
freedom~\cite{Bonini:1999cn,Bonini:1999kj}.  Hence, our strategy is
to evaluate numerically $F(\tilde{E},\tilde{N})$ in as large region of
the $(E,N)$ plane as possible, and then extrapolate the results to
$\tilde{N}\to0$.  In what follows we omit tilde over $E$ and $N$ to
simplify notations.

\begin{figure}
  \begin{center}%
    \includegraphics{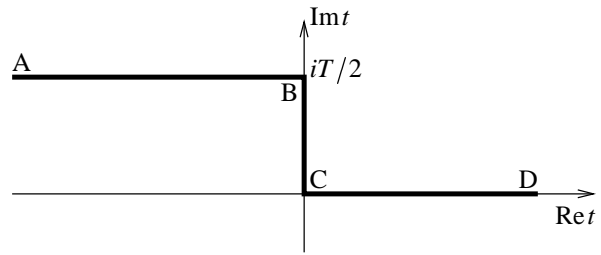}
  \end{center}
  \caption{The contour in complex time used in the formulation
    of the boundary value problem~\eqref{final_BVP}.}
  \label{fig:time_contour}
\end{figure}

The boundary value problem for $F(E,N)$ was derived
elsewhere~\cite{Rubakov:1992fb,Tinyakov:1992fn,Rubakov:1992ec}, so we
only present its formulation.  Let $\varphi(\bx,t)$ denote
collectively all physical fields in the model.  One introduces two
auxiliary real parameters $T$ and $\theta$ and considers
$\varphi(\bx,t)$ as complex functions on the contour ABCD 
in the complex time plane shown in
Fig.~\ref{fig:time_contour}.  The parameter
$T$ determines the height of the contour (equal to $T/2$),
while the role of $\theta$ will be described later.  
The field $\varphi$ should satisfy
the field equations,
\begin{subequations}\label{final_BVP}
\begin{equation}
  \label{BVP_eq}
  \vpar{S}{\varphi} = 0
\end{equation}
on the contour ABCD.  In the infinite future (part D of the contour), the
field should be real
\begin{equation}\label{BVP_real}
  \Imag \dot \varphi(\bm{x},T_f\to\infty) \to 0 \;,\quad
  \Imag \varphi(\bm{x},T_f\to\infty) \to 0
\end{equation}
(for complex fields, such as $\Phi$ in \eqref{SU2action}, this means
that both $(\Phi+\Phi^\dagger)/2$ and
$(\Phi-\Phi^\dagger)/2i$ must be real).
The remaining boundary conditions are imposed in the infinite past,
$t=iT/2+T_i$, $T_i\to-\infty$, part A of the contour.  Since for
$T_i\to-\infty$
the system reduces to a superposition
of non-interacting waves about one of the gauge theory vacua (which we
choose to be the trivial one for definiteness), the field $\varphi$
linearizes
\begin{align*}
  \varphi({\bm x},t)\big|_{t\to-\infty+iT/2} = \\
  \int \frac{d\bk}{ (2\pi)^{3/2}{\sqrt{2\omega_{\bk}}} }
   \Big( & \fk \e^{-i\omega_\bk(t-iT/2)+i{\bm{kx}}} \nonumber\\
         & + g_{\bk}^* \e^{i\omega_\bk(t-iT/2)-i{\bm{kx}}}  \Big)
  \;. \nonumber
\end{align*}
The boundary condition in the infinite past is then
the ``$\theta$ boundary condition''
\begin{equation}\label{BVP_bc2}
  \fk = \e^{-\theta} g_{\bk} \;.
\end{equation}
\end{subequations}
For $\theta$ different from zero this equation implies that the fields
themselves must be continued to complex values.  
For a complex field, like $\Phi$
in~\eqref{SU2action}, its real and imaginary parts must be continued to
complex values separately.  Finally, there are two more equations,
\begin{equation*}
  E = \int d\bk\, \omega\bkind \fk\gsk \;,\quad
  N = \int d\bk\, \fk\gsk \;.
\end{equation*}
These equations indirectly fix the values of $T$ and $\theta$ for
given energy and number of incoming particles.  Note that they
are in fact semiclassical
expressions for $E$ and $N$ in terms of the frequency components of
the incoming field.

Given a solution to the boundary value problem, the exponent for the
inclusive transition probability is
\begin{equation}\label{sigmaFexp}
  \frac{4\pi}{\alpha_W} F(E,N) =
  2\Imag S_{ABCD}(\varphi) - N\theta - ET
  \;.
\end{equation}
From Eq.~\eqref{sigmaFexp} the variables $(T,\theta)$
appear to be Legendre conjugates  to $(E,N)$.  This correspondence
is strengthened by the following relations,
\begin{eqnarray}
  \frac{4\pi}{\alpha_W} \frac{\partial F(E,N)}{\partial E} = -T
  \label{7a*} \\
  \frac{4\pi}{\alpha_W} \frac{\partial F(E,N)}{\partial N} = -\theta
  \;.\label{7a**}
\end{eqnarray}
These relations are useful as a cross check of the numerical
procedure,
and also as a mean of extrapolating $F(E,N)$ to $N=0$.

This method of obtaining the exponent for tunneling probability was
implemented in quantum mechanics of two degrees of
freedom~\cite{Bonini:1999cn,Bonini:1999kj,Bezrukov:2003yf} and in
scalar theory exhibiting collision-induced false vacuum
decay~\cite{Kuznetsov:1997az}.  It has been adapted to systems with
gauge degrees of freedom in Ref.~\cite{Bezrukov:2001dg} where
preliminary study of the energy region below $E_\mathrm{sph}$ was
performed.

Two remarks are in order.  First, the boundary value
problem~\eqref{final_BVP} by itself does not guarantee that its
solution interpolates between topologically distinct vacua.  Ensuring
that the solutions have correct topology is an independent and
important part of the computational procedure.

Second, we look for solutions to the boundary value
problem~\eqref{final_BVP}, which are \emph{spherically symmetric} in
space.  Physically, since both instanton and sphaleron have this property,
it is likely that the relevant solutions are also spherically
symmetric.  Technically, spherical symmetry
reduces the number of equations considerably, so that the numerical
analysis simplifies significantly.  In the gauge $A_0=0$, spherically
symmetric configurations~\cite{Ratra:1988dp} are parameterized by five
two-dimensional fields $a$, $\alpha$, $\beta$, $\mu$ and $\nu$,
\begin{align}
  A_i(\bm{x},t) &= \half\bigg[a_1(r,t)\bm{\sigma\cdot n}n_i
                   +\frac{\alpha(r,t)}{r}(\sigma_i
                                          -\bm{\sigma\cdot n}n_i)
  \nonumber\\
                &\hphantom{=\half\bigg[}
                   +\frac{1+\beta(r,t)}{r}\epsilon_{ijk}n_j\sigma_k
                   \bigg]
  \label{anzats}\\
  \Phi(\bm{x},t) &= [\mu(r,t)+i\nu(r,t)\bm{\sigma\cdot n}]\xi \;,
  \nonumber
\end{align}
where $\xi$ is a unit two-column.
The fields $a$, $\alpha$, $\beta$, $\mu$, $\nu$ are real in the
original $SU(2)$-Higgs theory, but they become complexified due to the
$\theta$-boundary condition~\eqref{BVP_bc2}.

We solved the boundary value problem~\eqref{final_BVP} 
numerically in the $A_0=0$ gauge on a grid of
spatial size in radial direction
$R=8/(\sqrt{2}M_W)$ and number of spatial grid points $N_r=90$.  The
length of initial Minkowskian part of the contour AB was equal to
$6/(\sqrt{2}M_W)$.  The number of time grid points on this part was
$N_t=200$ while on the Euclidean part $BC$ it was equal to 150.  The
number of time grid points on the part CD varied, with the maximum
number of about 400.

\begin{figure}
  \centerline{\includegraphics[width=\columnwidth]{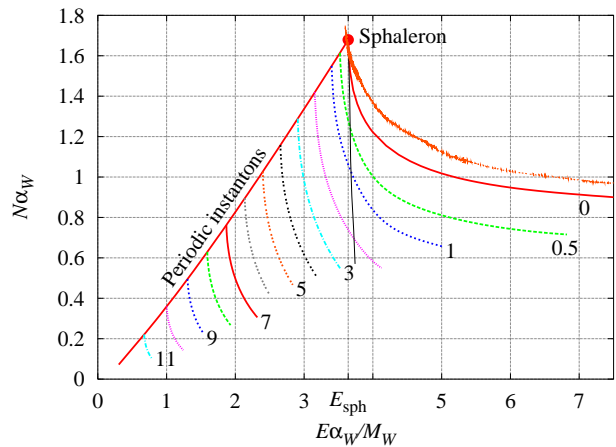}}
  \caption{Lines of $F(E,N)=\const$.  The lines are labeled by the values
    of the suppression exponent $-\alpha_W\log\sigma=4\pi F$.
    The line labeled by $0$ ($F=0$) is the boundary of
    the classically allowed region. The ``fuzzy'' line
    represents the approximate boundary of the classically allowed
    region found in the over-barrier calculations of
    Ref.~\cite{Rebbi:1996zx}.}
  \label{fig:ne_f}
\end{figure}

\begin{figure}
  \begin{center}
    \includegraphics[width=\columnwidth]{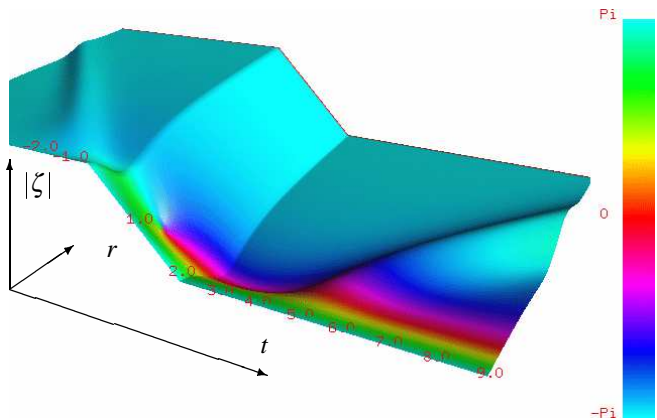}%
  \end{center}
  \caption{The field $\zeta$ for a solution with $N=1/\alpha_W$ and
    $E=3.35M_W/\alpha_W$.  The color tracks the phase of the field,
    whose behavior indeed shows the correct topology.  The part
    corresponding to the Euclidean evolution is inclined for
    visualization purposes.}
  \label{fig:surface_good}
\end{figure}

\begin{figure}
  \begin{center}
    \includegraphics[width=\columnwidth]{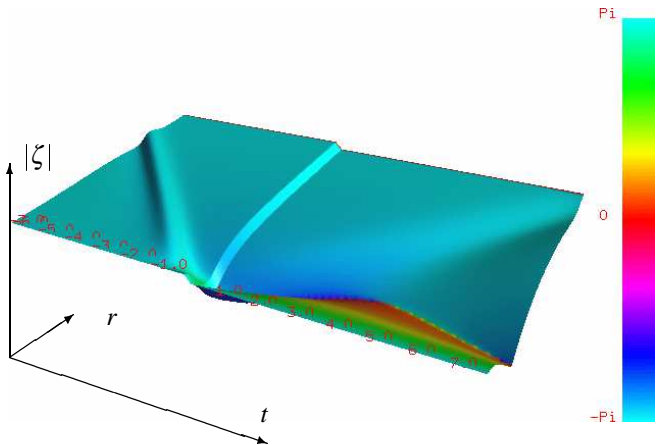}%
  \end{center}
  \caption{Field $\zeta$ for a solution with $N=1/\alpha_W$ and
  $E=4.48M_W/\alpha_W$.}
  \label{fig:surface_sph}
\end{figure}

The details of our numerical procedure are given
elsewhere~\cite{Bezrukov:2003er}. Here we concentrate on our results.
Clearly, only a part of $(E,N)$ plane is accessible to numerical study:
the difficulty of the calculation increases 
at higher energies and smaller number of
particles, as the solutions get sharper and linearize slower at large
negative times.
The region of $(E,N)$ plane covered in our study is shown in
Fig.~\ref{fig:ne_f}, where we present the results for $F(E,N)$.
Before discussing the tunneling exponent $F(E,N)$, let us comment on
various types of solutions we have found.  The
upper left line is the line of periodic instantons.  These are
solutions to the boundary value problem with
$\theta=0$~\cite{Khlebnikov:1991th,Bonini:1999fc}, which correspond to
transitions with the smallest tunneling exponent $F$ for given
energy.
The line of periodic instantons ends at the sphaleron
point\footnote{The number of incoming particles for the sphaleron is
obtained by infinitesimally perturbing the (unstable) sphaleron
solution along the negative mode, and integrating backwards in real
time~\cite{Rebbi:1996zx}.}.  The almost vertical line beginning at the
sphaleron in Fig.~\ref{fig:ne_f} separates two parts of the
classically forbidden region in which the solutions have qualitatively
different properties.  To the left of this line, the solutions are
real on the Minkowskian part CD of the contour, and rapidly dissipate
at large times forming spherical waves.  This is illustrated in
Fig.~\ref{fig:surface_good} where the field $\zeta=\beta-i\alpha$ is
shown for a solution with relatively low energy\footnote{Note that
$\beta=-1$, $\alpha=0$ corresponds to the trivial gauge theory vacuum,
while in the first topological vacuum $\zeta$ winds around the unit
circle in complex plane as $r$ runs from $0$ to $\infty$, see
Eq.~\ref{anzats}}.  On the other hand, in the right part of the
forbidden region, the solutions are complex on the part CD of the
contour, and obey the reality condition~\eqref{BVP_real} only
asymptotically.  Part of the energy is emitted away in the form of
spherical waves, but there remains a lump of energy near the origin,
as illustrated in Fig.~\ref{fig:surface_sph}.  We have checked that
this remaining lump is nothing but the sphaleron\footnote{The very fact that
the solution is real only asymptotically in time is due to the
existence of the unstable mode about the sphaleron.}.

The physical interpretation of the two types of solutions is as
follows.  At low energies, the system tunnels directly to the
neighboring gauge theory vacuum plus linear excitations above it.  At
energies higher than the sphaleron energy (precisely, to the right of
the almost vertical line in Fig.~\ref{fig:ne_f}), the system ends up
close to the sphaleron, with extra outgoing waves in the sphaleron
background.  In the latter case the system jumps on top of the
barrier.  This process is not precisely what is usually meant by
tunneling; still it occurs with exponentially small probability which
may be attributed to the rearrangement of the system from a collection
of highly energetic incoming waves to the soft lump of the fields given
by the sphaleron.
The method of obtaining solutions of the second type
was proposed in Ref.~\cite{Bezrukov:2003yf}, and we make use of this
method in our work (see Ref.~\cite{Bezrukov:2003er} for details).

Let us now concentrate on the results for the tunneling exponent
$F(E,N)$.
Our data are in agreement with analytical results for $F(E,N)$ in the
low energy region, see Ref.~\cite{Bezrukov:2003zn} for details.
Another interesting comparison can be made with the results of
Ref~\cite{Rebbi:1996zx}, where a Monte-Carlo technique was used
to find real-time overbarrier solutions
close to the boundary of the classically allowed region.
This technique produced an approximation (and, at
the same time, an upper bound) for the boundary of the classically
allowed region.  It is seen that the results of
Ref.~\cite{Rebbi:1996zx} are reasonably close to the boundary of the
classically allowed region found in our calculations.

To get insight into the suppression factor $F_{HG}(E)$ for actual
particle collisions, we have to extrapolate our data to $N=0$, see
Eq.~\eqref{5*}.  We present here two types of extrapolation.  The
first one produces \emph{lower bounds} on the suppression exponent
$F_{HG}(E)$ itself, while the second one gives an estimate for $F_{HG}(E)$.  
While the latter extrapolation has stronger predictive power at relatively
low energies, $E\lesssim2E_\mathrm{sph}$, the former extends to much
higher energies, so the two are complementary.

\begin{figure}
  \centerline{\includegraphics[width=\columnwidth]{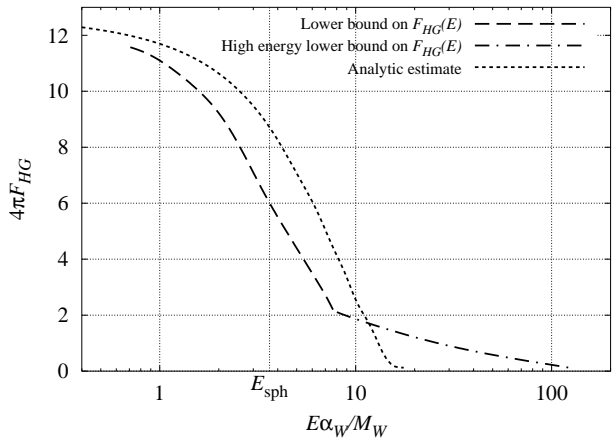}}
  \caption{Lower bounds on the suppression exponent for two-particle
    collisions, dashed and dashed-dotted lines.  The dotted line is the
    estimate of Refs.~\cite{Ringwald:2002sw,Ringwald:2003px}.}
  \label{fig:hg}
\end{figure}

\begin{figure}
  \centerline{\includegraphics[width=\columnwidth]{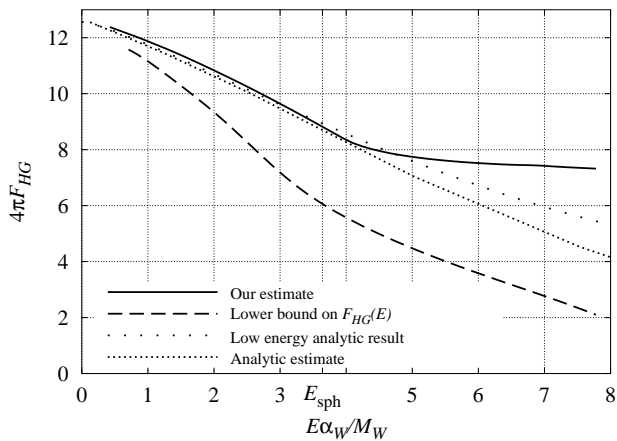}}
  \caption{Estimate of the suppression exponent for two-particle
    collisions $F_{HG}(E)$ (solid line), lower bound on $F_{HG}(E)$
    (dashed line), low energy analytic prediction~\eqref{HGanalytic}
    (rare dotted line) and analytic estimate of
    Refs.~\cite{Ringwald:2002sw,Ringwald:2003px} (dotted line).}
  \label{fig:hg_low}
\end{figure}

We begin with the lower bounds on $F_{HG}(E)$.  One way to obtain a
lower bound is to make use of Eq.~\eqref{7a**}, together with the fact
that $\theta$ increases as $N$ gets smaller.  Hence, a lower bound on
$F_{HG}(E)$ is obtained by simply continuing $F(E,N)$ with a linear
function of $N$ for each energy.  This bound is shown in
Figs.~\ref{fig:hg}, \ref{fig:hg_low}, dashed line.  It indicates that
up to the energy $8M_W/\alpha_W\simeq 20\TeV$ the suppression is still
high: the suppression factor is smaller than $\e^{-60}\sim10^{-26}$
for $\alpha_W\simeq1/30$. 

Another lower bound, the best we can obtain at very high energies, is
constructed by exploiting the observation that the lines of constant
$F$ in $(E,N)$ plane have positive curvature (see
Fig.~\ref{fig:ne_f}).  So, the lower bound is obtained by
extrapolating these lines linearly to $N=0$.  This bound is displayed
in Fig.~\ref{fig:hg}, dashed-dotted line.  One can see that
exponential suppression continues up to the energy of at least
$250~\TeV$.

Let us now come to the second type of extrapolation which we make to
estimate $F_{HG}(E)$ itself.  We find it appropriate to use
Eq.~\eqref{7a*}.  The point is that the function $T(N)$ at fixed
energy is approximately linear in $N$.  This property has been shown
analytically for low energies~\cite{Bezrukov:2003zn}, while for all
energies it follows from our numerical data.
[This is in contrast to the behavior of $F(E,N)$: the
analytical results at low energy show that for fixed $E$, this
function behaves as $N\log N$ as $N\to0$~\cite{Bezrukov:2003zn}.]  We
thus extrapolate $T(N,E$) linearly to $N=0$ along $E=\const$, and then
integrate Eq.~\eqref{7a*} at $N=0$ to obtain the suppression exponent
$F_{HG}(E)$ for two-particle collisions.  This estimate is shown in
Fig.~\ref{fig:hg_low}, solid line.  It is instructive to compare it to
the one loop analytic
result~%
\cite{Khoze:1991bm,Arnold:1991cx,Diakonov:LINPSchool1991,Mueller:1991fa},
which gives three terms in the low-energy expansion,
\begin{equation}\label{HGanalytic}
  \frac{4\pi}{\alpha_W}F(E) = \frac{4\pi}{\alpha_W}\left[
    1-\frac{9}{8}\left(\frac{E}{E_0}\right)^{4/3}
    +\frac{9}{16}\left(\frac{E}{E_0}\right)^2
  \right],
\end{equation}
where $E_0=\sqrt{6}\pi M_W/\alpha_W$.  We see that our numerical data
are (somewhat unexpectedly) very close to the one loop
result~\eqref{HGanalytic} up to the sphaleron energy.  In this energy
region, they are consistent also with the analytic estimate of
Refs.~\cite{Ringwald:2002sw,Ringwald:2003px}.  On the other hand, the
behavior of $F_{HG}(E)$ changes dramatically at $E\gtrsim
E_\mathrm{sph}$.  We attribute this to the change in the tunneling
behavior---at $E\gtrsim E_\mathrm{sph}$ the system tunnels ``on top of
the barrier''.  Our numerical data show that the suppression exponent
$F_{HG}(E)$ flattens out, and topology changing processes are in fact
much heavier suppressed at $E\gtrsim E_\mathrm{sph}$ as compared to
the estimate~\eqref{HGanalytic} and the estimate of
Refs.~\cite{Ringwald:2002sw,Ringwald:2003px}.

Thus, our numerical results, albeit covering a limited range of
energies and initial particle numbers, enable us to obtain both a lower
bound for and an actual estimate of the suppression exponent for the
topology changing two-particle cross-section in the electroweak theory
well above the sphaleron energy.  This cross section remains
exponentially suppressed up to very high energies of at least
$250~\TeV$.  In fact, the energy, if any, at which the exponential
suppression disappears, is most likely much higher, as suggested by
comparison of our lower bound and actual estimate at energies
exceeding significantly $E_\mathrm{sph}$, see Fig.~\ref{fig:hg_low}.

\begin{acknowledgments}
The authors are indebted to A.~Kuznetsov for helpful discussions.  We
wish to thank Boston University's Center for Computational Science and
Office of Information Technology for generous allocations of
supercomputer time.  Part of this work was completed during visits by
C.R.~at the Institute for Nuclear Research of the Russian Academy of
Sciences and by F.B., D.L., V.R. and P.T. at Boston University, and we
all gratefully acknowledge the warm hospitality extended to us by the
hosting institutions.  The work of PT was supported in part by the
SNSF grant 21-58947.99.  This research was supported by Russian
Foundation for Basic Research grant 02-02-17398, U.S.  Civilian
Research and Development Foundation for Independent States of
FSU~(CRDF) award RP1-2364-MO-02, and DOE grant US DE-FG02-91ER40676.
\end{acknowledgments}


\end{document}